\documentclass[reqno]{amsart}
\usepackage{graphics,amsmath}
\usepackage{tikz}
\usepackage{color}
\usetikzlibrary{shapes,arrows}

\newcount\n  \newdimen\w
  
\def\eqn#1#2{ \begin{align} \label{#1}         #2 \end{align}}

\def\nl#1{          \\ \label{#1}        }  
\def\nnl#1{ \tag*{} \\ \label{#1}        }  

\def\re#1{(\ref{#1})}   

\def\delim#1#2#3{\csname\ifcase#1 relax\or   
   big\or Big\or bigg\or Bigg\fi\endcsname   
  {\ifcase#2\or\Delim#3\or\deliM#3\fi}}      
\def\Delim#1{\ifcase#1\relax\or(\or[\or\{\or<\or\langle\or|\or\|\or---{ }\fi}
\def\deliM#1{\ifcase#1\relax\or)\or]\or\}\or>\or\rangle\or|\or\|\or{ }---\fi}

\let\f\frac                     

\begin{document}

\title{Thermodynamic Approach to Generalized Continua}
\author{P. V\'an$^{1,2,3}$, A. Berezovski$^4$ and C. Papenfuss$^{5}$}
\address{$^1$Dept. of Theoretical Physics, Wigner RCP, HAS, \\  H-1525 Budapest, P.O.Box 49, Hungary;
and  {$^2$Dept. of Energy Engineering, Budapest Univ. of Technology and Economics},\\
  H-1111, Budapest, Bertalan Lajos u. 4-6,  Hungary \\
$^3$Montavid Thermodynamic Research Group \\
$^4$Centre for Nonlinear Studies, Institute of Cybernetics at Tallinn Technical University, Akadeemia tee 21, 12618 Tallinn, Estonia \\
$^5$Technical University of Berlin, Strasse des 17. Juni 135, 10623 Berlin, Germany}
\email{Van.Peter@wigner.mta.hu}

\date{\today}
\begin{abstract}
 Governing equations of dissipative generalized solid mechanics are derived by thermodynamic methods in  the Piola-Kirchhoff framework using  the Liu procedure. The isotropic small strain case is investigated in more detail. The connection to  the Ginzburg-Landau type evolution, dual internal variables, and a thermodynamic generalization of  the standard linear solid model of rheology  is demonstrated. Specific examples are chosen to emphasize experimental confirmations and predictions beyond  less general approaches.
\end{abstract}
\maketitle

\section{Introduction}

Conventional theories of continua do not provide the description of a microstructural influence because material elements are considered as indistinct pieces of matter.
Generalized continuum theories (higher-order or higher-grade) are first examples of what has been proposed to describe the macroscopic behavior of materials with inner material structure.
Though their constitutive structure is restricted by the second law of thermodynamics,
the contribution of {small}-scale events to entropy fluxes and sources is still not completely investigated.

Governing equations of generalized continuum mechanics can be obtained by different  ways.
{The four} most widely accepted approaches {are the following}: the variational approach by Mindlin \cite{Min64a}, the microhomogenization procedure by Eringen an Suhubi \cite{EriSuh64a}, the virtual power method by Germain \cite{Ger73a} {and the Galiean invariance based considerations of Eringen \cite{Eri92a,Eri99b}}.

Mindlin \cite{Min64a} derived  governing equations in the small strain approximation with the help of a variational principle analogous to that in ideal elasticity.
In his theory of microdeformations, Mindlin introduced  kinetic and  potential energy  for both  micro- an macro-displacements {as well as} a tensor characterizing micro-inertia.
In the Mindlin theory, the  potential energy density is a quadratic function of the macro-strain, the relative strain, and the micro-deformation gradient.
With this variational foundation, the Mindlin theory is an idealized one, which does not include {any} dissipation.

In their approach to generalized continua, Eringen and Suhubi \cite{EriSuh64a,Eri99b} did not start from a variational principle.
They obtained  an evolution equation of the micro-strain extending  mechanical concepts of inertia, stress, strain, and energy onto the microlevel and calculating velocity moments of the microevolution of the momentum.
The zeroth moment of the mesoscopic momentum balance results in the macroscopic Cauchy equation and the first moment gives the evolution equation of the micromomentum.
In their constitutive theory, Eringen and Suhubi consider the microdeformation and the gradient of the microdeformation as general internal variables and calculate the entropy production accordingly.
They obtained the following expression for the entropy production $ T\Delta$(\cite{EriSuh64a} (5.13)):
\eqn{ep1}
{
\begin{split}
  T\Delta =  & q_i \partial_i \log{T} + (\tau_{ij} - \tilde \tau_{ij})\partial_iv_j +
  (s_{kl} - \tau_{kl} - \tilde \sigma_{kl})\nu_{kl} + \\
   + &  (\mu_{klm} - \tilde \mu_{klm}) \partial_{m}\nu_{kl} - \rho \zeta_a\dot \xi_a \geq 0.
\end{split}
}
Here $q_i$ is the heat flux,  {i.e. flux} of the internal energy density, $T$ is the temperature, $\tau_{ij}$ is the stress, {$\rho$ is the matter density,}
$s_{ij}$ is the relative stress,  
$\nu_{ij}$ is the microvelocity gradient, $\mu_{ijk}$ is the double stress, 
the derivative of the free energy with respect to the microdeformation gradient.
Tilde denotes the reversible, nondissipative parts of the corresponding quantities.
{Further,} $\tilde \sigma_{ij}= \tilde t_{ij} - \tilde s_{ij}$ is the reversible microstress, the derivative of the free energy with respect to the microdeformation.
{Equation \re{ep1}} defines $\tilde s_{ij}$, the  reversible relative stress, too.
The last term gives the entropy production due to additional internal variables $\xi_a$, $a=1,..n$, {and} $\zeta_a$ denotes the related intensive variable, the derivative of the entropy density with respect to $\xi_a$.
Here the notation of derivatives and indices follow Mindlin as far as it was possible.
Remarkable is that the Eringen-Suhubi theory defines the entropy {flux} in its classical form $J_i = q_i/T$.

{The second method of Eringen starts from the energy balance and derives the additional balances from the requirement of Galiean invariance \cite{Eri92a,Eri99b,LeeWan11a}. However, the final equations and the consitutive theory is identical with the preivous case.}

As we can see, the descriptions of generalized continua are weakly nonlocal from the beginning because  the gradient of the microdeformation is introduced as {a} state variable.
This is true also for the third approach that uses the principle of virtual power in order to derive the evolution equation of the microdeformation.
The virtual power method is essentially dealing with statics.
 Dynamics in this method is introduced by  {an} assumption that inertia is connected to virtual displacements directly \cite{Ger73a,ForEta08a} or with the help of  dissipation potentials.
In the first case, the  dissipation requires  separate considerations.
In the second case, the  dissipation  is introduced together with inertial terms.

Let us underline some common properties of  above mentioned {approaches}:
\begin{enumerate}
 \item \textit{The connection of new microstructural variables to mechanical effects has a kinematic background.} However, the microdeformation can be originated in different structural changes (e.g. microcracking), which is not necessarily connected to energy alterations due to the change of a Riemann geometry of the material manifold.
 \item \textit{The evolution equations of  microstructural variables are originated in mechanics.} Variational principles, {moment series developement as well as} 
 virtual power with dissipation potentials are mechanical concepts. {In the derivation of the governing equations the} dissipative effects are considered as secondary and frequently disregarded.
 \item \textit{The entropy flux has the same form} as that in the simplest case of Cauchy continua.
 \item In all these approaches, it is customary to introduce internal variables independently of the microdeformation, representing already identified structural changes of continua, e.g. damage or cracking \cite{AslFor11c,PapFor06a}.
\end{enumerate}
Intuitively, it is natural to expect that the microstructure affects also dissipative phenomena.
It is needed, therefore, to investigate in detail how dissipation effects are described by generalized continuum theories.

The response of materials to external loads can be expressed explicitly as a functional \cite{Hau93a}. On the other hand, additional internal variables can be introduced to define this functional in an implicit manner by means of their evolution equations.

It is well known, that variational principles may exist for a dissipative evolution, but they are not of the usual Hamiltonian kind.
At best, they need to be modified because they  do not work without any further ado \cite{Gya70b,VanMus94a,VanNyi99a}.
Regarding the homogenization technics of Eringen and Suhubi for internal variables \cite{EriSuh64a}, it should be noted that  there is no primary microscopic candidate for an evolution equation, therefore their method cannot be applied beyond the kinematic determinations.
Direct statistical or kinetic theory related calculations (as for example \cite{IrvKir50a}) would require a particular microstructure and an interpretation of the internal variable. {The original approach of mesoscopic theory, motivated by mixtures, does not require a very detailed mesoscopic interpretation, but the treatment of dissipative phenomena is not complete there \cite{BleMus91a,EhrAta97a}}
%
%
Finally, regarding the third treatment, the principle of virtual power is a mechanical concept, and internal variables {(if any)} are not related to any kind of spacial changes of continua.

It is known, that there are two alternate approaches to obtain the evolution equations of internal variables directly. The thermodynamic approach is related to {\em internal variables of state} and the variational one is regarded to {\em dynamic degrees of freedom}, respectively \cite{MauMus94a1,MauMus94a2}.
In the thermodynamic approach,  the entropy production is calculated considering  additional constraints (e.g. balances), and the evolution of  internal variables is determined as a part of the constitutive theory by means of the {dissipation} inequality.
Evolution equations  obtained {in such a way} {contain} typically {only} first order time {derivatives} \cite{ColGur67a}. 
The most important issues here are the stability of the weakly nonlocal extension (see critical remarks in \cite{MulWei12a}), the thermodynamic consistency, and seemingly missing boundary conditions. In the case of dynamic degrees of freedom, a Hamiltonian variational principle is applied to the nondissipative part of the evolution, and the dissipative contribution is calculated by dissipation potentials.

Recently, it was shown that one can get a unified description of the two 
methods introducing weakly nonlocal dual internal variables \cite{VanAta08a}.
Then the exact and constructive exploitation of the entropy inequality, e.g. the Liu procedure, combined with an Onsagerian linear approximation of constitutive functions leads to completely solvable constraints.
An essential ingredient of the approach of dual internal variables is the observation that in the case of higher-order gradient theories,  gradients of constraints of the entropy inequality are constraints on the constitutive state space \cite{Van04a,VanFul06a,Cim07a}.

It has been observed recently \cite{BerEta11a} that the structure of the dual internal variable system of evolution equations in the nondissipative case corresponds exactly to the evolution equation in the Mindlin theory. In this case, one of the internal variables can be interpreted as the microdeformation and another  as the conjugated momentum.
It is worth to extend the dual internal variable approach to a broader class of materials taking into account dissipative effects.
This suggests a more general procedure to construct  evolution equations than in \cite{VanAta08a}.
The corresponding procedure is presented in the paper.
It consists in the extension of the state space, the formulation of constraints, the application of the Liu procedure to the entropy inequality,
the solution of obtained  Liu equations, and the specification of a general form of  evolution equations for internal variables following from  linear relationships between thermodynamic fluxes and forces.
The relation to linear viscoelasticity, to the 
{pattern formation} equations of the Ginzburg-Landau-type, and to the standard linear solid model is demonstrated in the small-strain approximation.
The same approximation is used to point out the extension of generalized continua descriptions onto dissipative materials and microstructural thermal effects.

\section{Construction of evolution equations}

\subsection{{Balance laws}}

We start with the  formulation of thermodynamic constraints for continuum mechanics with dual internal variables in {the} Piola-Kirchhoff framework (PK frame). {In this way the following calculations are technically easier.}

The balances of momentum and energy {can be represented as follows:}
\eqn{mom_bal}{
\varrho_0 \dot{v}_i - \partial_j t_{ij} &= 0, \nl{inte_bal}
\varrho_0 \dot{e} + \partial_i q_i &= t_{ij} \partial_jv_i,
}
where \(\rho_0\) is the {matter} density, $v_i$ is the velocity field,  \(t_{ij}\) is the first Piola-Kirchhoff stress, \(e\) is the specific internal energy and \(q_i\) is the {flux}  of internal energy density in the PK frame.
The dot denotes the material time derivative, which is a partial time derivative on the material manifold.
$\partial_j$ denotes the (material) {space derivative}.
The Einstein summation rule  is applied for repeated indices.
Our index notation is abstract, does not refer to any particular system of coordinates, {and} denotes the tensorial degree and  contractions in accordance with the traditional coordinate free treatment in continuum mechanics \cite{Wal84b}.
In the notation we use uniformly lowercase indices, i.e. do not distinguish between material and spatial indices and vectors and covectors.
This way it is easier to follow calculations, and {examples  are considered in the small strain approximation} where the differences are {negligible}.

The balances are introduced without source terms, the momentum and the total energy {are} conserved, because  source terms are irrelevant in a constitutive theory.
The relation {between} the deformation gradient $F_{ij}$ and the velocity field is {considered as a} constraint:\
\eqn{kincon}{
\dot F_{ij} - \partial_jv_i =0.
}
Evolution equations of the internal variables \(\psi_{ij}\) and \(\beta_{ij}\) are {formally represented as}
\eqn{evolint}{
\dot{\psi}_{ij} + f_{ij}=0, \qquad \dot{\beta}_{ij} + g_{ij} = 0,
}
{where the constitutive functions $f_{ij}$ and $g_{ij}$ depend on the whole set of state variables}.
Here the notation of Mindlin was applied for the first internal variable, $\psi_{ij}$, which is the microdeformation there \cite{Min64a}.

The entropy inequality is given as follows:
\eqn{entr_b0}{
\varrho_0 \dot s + \partial_i J_i \geq 0,
}
where $s$ is the specific entropy and $J_i$ is the {flux of} the entropy density.

{It is assumed that} constitutive functions
\begin{equation}
t_{ij}, q_i, f_{ij}, g_{ij}, s, J_i,
\end{equation}
are defined on the weakly nonlocal  state space spanned by the following variables:
\begin{equation}
\partial_j v_i, F_{ij}, \partial_k F_{ij}, e, \partial_i e,\psi_{ij}, \partial_k \psi_{ij}, \partial_{kl}\psi_{ij}, \beta_{ij}, \partial_k \beta_{ij}, \partial_{kl} \beta_{ij}.
\end{equation}
For the sake of simplicity, we consider a weakly nonlocal constitutive state space of the first order  in the deformation gradient and in the internal energy, but of the second order in the internal variables, i.e., their second gradients are included.
The velocity field is distinguished, because only its derivative is present in the constitutive state space.
{This} assumption  allows us to avoid  velocity related problematic aspects of the  material frame indifference (see e.g. \cite{MatVan06a}) {and corresponds to acceleration insensitive materials}.

\subsection{Liu procedure}

Balance of linear momentum \re{mom_bal},  balance of internal energy \re{inte_bal},   kinematic {relation} \re{kincon},  and  evolution equations of  internal variables \re{evolint} are constraints of the entropy inequality.
Taking into account that the constitutive state space is  weakly nonlocal {of the} second order in the internal variables, we should introduce additional constraints
{for} derivatives of their evolution equations.
We consider a derivative of a constraint as a new constraint in case of higher order weakly nonlocal constitutive state space.
It is an important aspect for the development of correct thermodynamic conditions in weakly nonlocal thermodynamic theories \cite{Van05a,Van09a1}:
\eqn{devolint}{
\partial_k \dot{\psi}_{ij} +\partial_k f_{ij}=0, \qquad \partial_k \dot{\beta}_{ij} +\partial_k g_{ij}=0.
}
Then we  introduce  Lagrange-Farkas multipliers $\lambda_i, \kappa, \Lambda_i, A_{ij}, \mathcal{A}_{ijk}, B_{ij}, \mathcal{B}_{ijk}$ for  constraints \re{mom_bal}, \re{inte_bal}, \re{kincon}, \re{evolint}$_1$, \re{devolint}$_1$, \re{evolint}$_2$, \re{devolint}$_2$, respectively.
The constrained entropy {im}balance is, {therefore,}
\eqn{conentrbal}{
\begin{split}
&\varrho_0 \dot{s} + \partial_j J_j -
  \Lambda_{ji} \left(\dot{F}_{ij} - \partial_jv_i  \right) -
  \lambda_i \left( \varrho_0 \dot{v}_i - \partial_j t_{ij} \right) -
  \kappa \left( \varrho_0 \dot{e} + \partial_i q_i  - t_{ij} \partial_j v_i  \right) + \\
& +A_{ji} \left( \dot{\psi}_{ij} + f_{ij} \right) +
  \mathcal{A}_{kji} \left( \partial_k \dot{\psi}_{ij} + \partial_k f_{ij} \right) + \\
& + B_{ji} \left( \dot{\beta}_{ij} + g_{ij} \right) +
  \mathcal{B}_{kji}  \left(\partial_k \dot{\beta}_{ij} + \partial_k g_{ij} \right)\geq 0.
  \end{split}
}
 Liu equations are obtained as coefficients of higher derivatives after a straightforward calculation:
\eqn{L0}{
\dot{v}_i\ :& \quad 0  =  \lambda_i \nl{L1}
\partial_j \dot{v}_i\ :& \quad \partial_{\partial_jv_i} s  =  0  \nl{L2}
\dot{F}_{ij}\ :& \quad \varrho_0\ \partial_{F_{ij}} s = \Lambda_{ij}  \nl{L3}
\partial_k\dot{F}_{ij}\ :& \quad  \partial_{\partial_k F_{ij}} s = 0  \nl{L4}
\dot{e} \ :& \quad \partial_e s =\kappa  \nl{L5}
\partial_i \dot{e} \ :& \quad \partial_{\partial_i e} s =0  \nl{L6}
\dot{\psi}_{ij} \ :& \quad  \varrho_0\  \partial_{ \psi_{ij}} s = - {A}_{ji}   \nl{L7}
\partial_k\dot{\psi}_{ij} \ :& \quad  \varrho_0\  \partial_{\partial_k \psi_{ij}} s = - \mathcal{A}_{kji}   \nl{L8}
\partial_{kl}\dot{\psi}_{ij} \ :& \quad  \varrho_0\  \partial_{\partial_{kl} \psi_{ij}} s = 0    \nl{L9}
\dot{\beta}_{ij} \ :& \quad  \varrho_0\  \partial_{ \beta_{ij}} s = - {B}_{ji}   \nl{L10}
\partial_k\dot{\beta}_{ij} \ :& \quad  \varrho_0\  \partial_{\partial_k \beta_{ij}} s = - \mathcal{B}_{kji}   \nl{L11}
\partial_{kl}\dot{\beta}_{ij} \ :& \quad  \varrho_0\  \partial_{\partial_{kl} \beta_{ij}} s = 0 \nl{L13}
\partial_{kj} v_i\ :& \quad \partial_{\partial_k v_i} J_j  -  \partial_{e}s\ \partial_{\partial_kv_i } q_{j}
  -\varrho_0  \partial_{\partial_k\psi_{lm}}s\ \partial_{\partial_jv_i } f_{lm}- \nnl{L131} &\qquad
  \varrho_0 \partial_{\partial_k\beta_{lm}}s\ \partial_{\partial_jv_i } g_{lm} =0  \nl{L132}
\partial_{kl} F_{ij}\ :& \quad \partial_{\partial_kF_{ij}} J_l  -  \partial_{e}s\  				
  \partial_{\partial_kF_{ij}}\  q_{l} -
  \varrho_0\ \partial_{\partial_k\psi_{nm}}s \partial_{\partial_lF_{ij} }\ f_{nm}- \nnl{L141}
  &\qquad \varrho_0\ \partial_{\partial_k\beta_{nm}}s\ \partial_{\partial_lF_{ij} }\ g_{nm} =0  \nl{L14}
\partial_{ij} e\ :& \quad \partial_{\partial_i e} J_j -  \partial_{e}s\  \partial_{\partial_i e }\ q_{j} -
  \varrho_0\ \partial_{\partial_j\psi_{lm}}s\  \partial_{\partial_i e }\ f_{lm}- \nnl{L151} &\qquad
  \varrho_0\ \partial_{\partial_j\beta_{lm}}s\  \partial_{\partial_i e }\ g_{lm} =0  \nnl{L15}}

\eqn{L162}{
\partial_{klm} \psi_{ij}\ :& \quad \partial_{\partial_{lm}\psi_{ij}} J_k -  \partial_{e}s\  				  \partial_{\partial_{lm}\psi_{ij}}\ q_{k} -
  \varrho_0\ \partial_{\partial_k\psi_{op}}s\ \partial_{\partial_{lm}\psi_{ij}}\ f_{op}- \nnl{L161} & \qquad
  \varrho_0\ \partial_{\partial_k\beta_{op}}s\ \partial_{\partial_{lm}\psi_{ij}}\  g_{op} =0  \nl{L16}
\partial_{klm} \beta_{ij}\ :& \quad \partial_{\partial_{lm}\beta_{ij}}\
  J_k  -  \partial_{e}s\  \partial_{\partial_{lm}\beta_{ij}}\ q_{k} -
  \varrho_0\   \partial_{\partial_k\beta_{op}}s\  \partial_{\partial_{lm}\beta_{ij}}\ f_{op}- \nnl{L171} & \qquad
  \varrho_0\   \partial_{\partial_k\beta_{op}}s\  \partial_{\partial_{lm}\beta_{ij}}\ g_{op} =0
}
 Liu equations \re{L0}, \re{L2}, \re{L4}, \re{L6}, \re{L7}, \re{L9} and \re{L10} {determine}  Lagrange-Farkas multipliers by  {corresponding} entropy derivatives.
 {The} solution of Liu equations \re{L1}, \re{L3}, \re{L5}, \re{L8} and \re{L11} reduces the constitutive form of the entropy {to the following one:} $s=s(F_{ij}, e, \psi_{ij}, \partial_k \psi_{ij}, \beta_{ij}, \partial_k \beta_{ij})$.
 The dissipation inequality then follows considering Liu equations \re{L131}-\re{L171} for the entropy {flux}.
 Together with the mentioned form of the entropy function, these equations can be solved and we obtain the entropy {flux in the form}:
\eqn{entc}{
J_i = \partial_{e}s q_{i} +
  \varrho_0 \partial_{\partial_i\psi_{lm}}s\   f_{lm} +
  \varrho_0 \partial_{\partial_i\beta_{lm}}s\  g_{lm} + J_i^0.
}
Here {the dependence of} $J_i^0$ is reduced similarly to that of the specific entropy $J_i^0=J_i^0(F_{ij}, e, \psi_{ij}, \partial_k \psi_{ij}, \beta_{ij}, \partial_k \beta_{ij})$.
{It should be noted} that Eq. \re{entc} is not the most general solution of corresponding Liu equations, {because} we did not consider  symmetries of  
functions in Eqs. \re{L8}, \re{L11} and \re{L13}-\re{L171}.
For example, Eq. \re{L13} is obtained as multiplier of $\partial_{kj}v_i$, therefore only the symmetric part of Eq. \re{L13} must be zero, but  we consider, however, solutions that are more restrictive. {This generality is more than enough to derive dissipative generalization of generalized mechanics.}

The dissipation inequality then follows as
\eqn{dissineq}{
\partial_{k} F_{ij} \left(  \partial_{F_{ij}} J_k  -  \partial_{e}s\  \partial_{F_{ij} }\ q_{k}
  -\varrho_0\   \partial_{\partial_k\psi_{lm}}s\  \partial_{F_{ij} }\ f_{lm}- \varrho_0 \  \partial_{\partial_k\beta_{lm}}s\  \partial_{F_{ij} }\ g_{lm}\right) + \nnl{di1}
\partial_{i}e\  \left(  \partial_{e} J_i  -  \partial_{e}s\  \partial_{e }\ q_{i}  -
  \varrho_0   \partial_{\partial_i\psi_{lm}}s\  \partial_{e }\ f_{lm}- \varrho_0\   					\partial_{\partial_i\beta_{lm}}s\  \partial_e g_{lm}\right) +  \nnl{di2}
\partial_{k}\  \psi_{ij}\left(  \partial_{\psi_{ij}} J_k  -  \partial_{e}s\  \partial_{\psi_{ij} }\ q_{k} -
  \varrho_0   \partial_{\partial_k\psi_{lm}}s\  \partial_{\psi_{ij} }\ f_{lm}- \varrho_0 \  					\partial_{\partial_k\beta_{lm}}s\  \partial_{\psi_{ij} }\ g_{lm}\right) +\nnl{di3}
\partial_{lk} \psi_{ij}\left(  \partial_{\partial_l\psi_{ij}} J_k  -  \partial_{e}s\
  \partial_{\partial_l\psi_{ij} }\ q_{k} -
  \varrho_0\ \partial_{\partial_k\psi_{om}}s\ \partial_{\partial_l\psi_{ij} }\ f_{om}-
  \varrho_0\ \partial_{\partial_k\beta_{om}}s\  \partial_{\partial_l\psi_{ij} }\ g_{om}\right) +\nnl{di4}
\partial_{k} \beta_{ij}\left(  \partial_{\beta_{ij}} J_k  -
  \partial_{e}s\ \partial_{\beta_{ij} }\ q_{k} - \varrho_0\  \partial_{\partial_k\psi_{lm}}s\  \partial_{\beta_{ij} }\ f_{lm}-
  \varrho_0\ \partial_{\partial_k\beta_{lm}}s\ \partial_{\beta_{ij} }\ g_{lm}\right) +\nnl{di5}
\partial_{lk}\  \beta_{ij}\left(  \partial_{\partial_l\beta_{ij}} J_k -  \partial_{e}s\
  \partial_{\partial_l\beta_{ij} }\ q_{k}  - \varrho_0 \   \partial_{\partial_k\psi_{om}}s\  \partial_{\partial_l\beta_{ij} }\ f_{om}- \varrho_0\   \partial_{\partial_k\beta_{om}}s\  \partial_{\partial_l\beta_{ij} }\ g_{om}\right)+ \nnl{di6}
\varrho_0 \  \partial_{F_{ij}}s\ \dot F_{ij}  +\partial_e s\ t_{ij}\partial_iv_j -
  \varrho_0\   \partial_{\psi_{ij}}s\  f_{ij}- \varrho_0\   \partial_{\beta_{ij}}s\  g_{ij}\geq 0
}
By substituting Eq. \re{entc} into the dissipation inequality we arrive at the following expression:
\eqn{entrpr2}{
\begin{split}
   & \partial_k \left( \partial_e s  \right) q_k +
\partial_e s\ t_{ij} \partial_iv_j +
\partial_j v_i \varrho_0 \partial_{F_{ij}}s  - 
\\
    & -\left(   \partial_{\psi_{ij}}s - \partial_k  \partial_{\partial_k\psi_{ij}}s  \right)\varrho_0 f_{ij} - \left(\partial_{\beta_{ij}}s - \partial_k  \partial_{\partial_k\beta_{ij}}s  \right)\varrho_0 g_{ij} + \partial_i J^0_i \geq 0.
\end{split}
}
Here $\partial_es = 1/T$, and we may identify thermodynamic fluxes and forces {as follows:}
\eqn{entrpr3}{
\begin{split}
   \partial_i \left( \frac{1}{T} \right)  q_i & +
\frac{1}{T}\underbrace{\left(t_{ij} + \varrho_0 T \partial_{F_{ij}}s \right)}_{{t}^v_{ij}}\partial_i v_j - \underbrace{ \left(\partial_{\psi_{ij}}s - \partial_k \partial_{\partial_k\psi_{ij}}s \right)}_{-X_{ij}}\varrho_0 f_{ij}- 
\\
    & -\underbrace{ \left( \partial_{\beta_{ij}}s - \partial_k  \partial_{\partial_k\beta_{ij}}s  \right)}_{-Y_{ij}}\varrho_0 g_{ij} +
\partial_i J^0_i \geq 0,
\end{split}
}
where $t^v_{ij}$ is the viscous stress.
The entropy is a distinguished constitutive function, that fixes the static information of the system. All other constitutive functions are determined by means of entropy derivatives.

\subsection{{Evolution equations}}
The  entropy production in the dissipation inequality  is {represented as} a sum of products, and there is  an undetermined constitutive function in each term
multiplied by a given function of the constitutive state space.
Therefore, it is straightforward to {point out} the simplest solution of the {dissipation} inequality assuming linear relationships {between}  thermodynamic fluxes (terms with undetermined constitutive functions) and their multipliers, thermodynamic forces {(see Table 1)}.

\begin{center}
{\small
\begin{tabular}{c|c|c|c|c}
       & Thermal & Mechanical & Internal 1 & Internal 2 \\ \hline
Fluxes & $ q_i$ &
    $\left(  t_{ij} + T \varrho_0 \partial_{F_{ij}}s\right)/T$ &
    $f_{ij}$ &
    $g_{ij}$\\ \hline
Forces &$\partial_i \left( \frac{1}{T} \right)$ &
    $\partial_i v_j$ &
    $\varrho_0\!\left(\partial_{\psi_{ij}}s\! -\! \partial_k  \partial_{\partial_k\psi_{ij}}s\right)$ &
    $\varrho_0\!\left(\partial_{\beta_{ij}}s\! -\! \partial_k  \partial_{\partial_k\beta_{ij}}s \right)$\\
\end{tabular}\\
\vskip .21cm
}
{Table 1. Thermodynamic fluxes and forces}
\end{center}
The classical thermal interaction is vectorial, {while other terms are tensorial}.
The mechanical term is responsible for viscoelasticity if no other terms are present.
The last two terms constitutively determine  evolution equations of second-order tensorial internal variables $\psi_{ij}$ and $\beta_{ij}$.
In isotropic  materials \re{iso1}-\re{iso}, tensorial mechanical and internal variables may be coupled independently of the {vectorial} thermal constitutive function:
\eqn{Ons1}{
{q}_i &= \lambda \partial_i \frac{1}{T} \nl{Ons2}
\frac{1}{T}t^v_{ij} &= \frac{1}{T}\left(  t_{ij} + \varrho_0 T  \partial_{F_{ij}}s\right) &=
  L^{11}_{ijkl} \partial_k v_l + L^{12}_{ijkl} X_{kl} + L^{13}_{ijkl} Y_{kl} \nl{Ons3}
\dot{\psi}_{ij} &= f_{ij} &=
  L^{21}_{ijkl} \partial_k v_l + L^{22}_{ijkl} X_{kl} + L^{23}_{ijkl} Y_{kl} \nl{Ons4}
\dot{\beta}_{ij} &= g_{ij} &=
  L^{31}_{ijkl} \partial_k v_l + L^{32}_{ijkl} X_{kl} + L^{33}_{ijkl} Y_{kl}
}
{So we have }  the three times three matrix of fourth order tensors $L^{IJ}_{ijkl}$  called \textit{conductivity tensors}.
In isotropic  materials they  { are characterized by} three scalar material parameters {each} and can be {represented} for all $I,J = 1,2,3$ as follows:
\eqn{iso1}{
 L^{IJ}_{ijkl} = l^{IJ}_1 \delta_{ik}\delta_{jl} + l^{IJ}_2 \delta_{il}\delta_{jk} + l^{IJ}_3 \delta_{ij}\delta_{kl}.
}
where $\delta_{ij}$ is the Kronecker delta.
Therefore, there are 1+3x9 = 28 {material} conductivity coefficients in  isotropic media with dual internal variables.
{ Representation} \re{iso1} can be decomposed into traceless symmetric, antisymmetric (deviatoric) and spherical parts, { i.e., is  equivalent to}
\eqn{iso}{
 L^{IJ}_{ijkl} = s^{IJ} \delta_{i\langle k}\delta_{jl\rangle} + a^{IJ} \delta_{i[k}\delta_{jl]} + l^{IJ} \delta_{ij}\delta_{kl},
}
where  braces $\langle \rangle\ $ denote the traceless symmetric part of the corresponding tensor in  related indices
$\delta_{i\langle k}\delta_{jl\rangle} = (\delta_{ik}\delta_{jl}+\delta_{jk}\delta_{il})/2 -\delta_{ij}\delta_{kl}/3$
and the rectangular parenthesis $[\ ]$ denotes the antisymmetric part as
$\delta_{i[k}\delta_{jl]} = (\delta_{ik}\delta_{jl}-\delta_{jk}\delta_{il})/2$.
Therefore, $s^{IJ}=l^{IJ}_1+l^{IJ}_2$, $a^{IJ} = l^{IJ}_1-l^{IJ}_2$, and $l^{IJ}=(3 l^{IJ}_3 + l^{IJ}_1 + l^{IJ}_2)/3$.
This kind of decomposition is instructive because symmetric, antisymmetric, and spherical second-order tensors are mutually orthogonal in the ``double dot'' product, i.e. taking the trace of their product.
Therefore,  constitutive equations \re{Ons2}-\re{Ons4} can be decomposed into three parts:  five component traceless symmetric, three component antisymmetric, and one component spherical  parts are independent.
The spherical part is {determined as}
\eqn{Ons_1}{
t^v_{kk} &=
  l^{11}\partial_k v_k + l^{12} X_{kk} + l^{13} Y_{kk}, \nl{Ons_2}
\dot{\psi}_{kk} &=
  l^{21}\partial_k v_k + l^{22} X_{kk} + l^{23} Y_{kk}, \nl{Ons_3}
\dot{\beta}_{kk} &=
  l^{31}\partial_k v_k + l^{32} X_{kk} + l^{33} Y_{kk}.
}
{In its turn, for} the symmetric traceless, deviatoric part {we have}
\eqn{Ons_s1}{
t^v_{\langle ij \rangle} &=
  s^{11} \partial_{\langle i} v_{l\rangle} + s^{12} X_{\langle ij \rangle} + s^{13}Y_{\langle ij \rangle}, \nl{Ons_s2}
\dot{\psi}_{\langle ij \rangle} &=
  s^{21} \partial_{\langle i} v_{l\rangle} + s^{22} X_{\langle ij \rangle} + s^{23}Y_{\langle ij \rangle}, \nl{Ons_s3}
\dot{\beta}_{\langle ij \rangle} &=
 s^{31} \partial_{\langle i} v_{l\rangle} + s^{32} X_{\langle ij \rangle} + s^{33}Y_{\langle ij \rangle}
}
and the antisymmetric part is given in a tensorial form instead of the vectorial invariants:
\eqn{Ons_a1}{
t^v_{[ij]} &=
  a^{11} \partial_{[i} v_{l]} + a^{12} X_{[ij]} + a^{13}Y_{[ij]}, \nl{Ons_a2}
\dot{\psi}_{[ij]} &=
  a^{21} \partial_{[i} v_{l]} + a^{22} X_{[ij]} + a^{23}Y_{[ij]}, \nl{Ons_a3}
\dot{\beta}_{[ij]} &=
 a^{31} \partial_{[i} v_{l]} + a^{32} X_{[ij]} + a^{33}Y_{[ij]}.
}
Therefore, conductivity matrices of spherical, traceless symmetric, and antisymmetric components are
\eqn{cmat}{
l^{IJ} =
\begin{pmatrix}
l^{11} & l^{12} & l^{13}\\
l^{21} & l^{22} & l^{23}\\
l^{31} & l^{32} & l^{33}
\end{pmatrix}, \quad
s^{IJ} =
\begin{pmatrix}
s^{11} & s^{12} & s^{13}\\
s^{21} & s^{22} & s^{23}\\
s^{31} & s^{32} & s^{33}
\end{pmatrix}, \quad
a^{IJ} =
\begin{pmatrix}
a^{11} & a^{12} & a^{13}\\
a^{21} & a^{22} & a^{23}\\
a^{31} & a^{32} & a^{33}
\end{pmatrix}.
}

The second law requires that  symmetric parts of 3x3 conductivity matrices $l^{IJ}$, $s^{IJ}$, and $a^{IJ}$  are positive definite.

With this requirement the thermodynamically consistent construction of continuum mechanical theory with second order tensorial dual internal variables is complete. Balances of momentum and energy  \re{mom_bal}, \re{inte_bal}  and evolution equations of the internal variables \re{evolint} together with  isotropic constitutive functions \re{Ons1}-\re{Ons4} form a complete system, if a particular form of the entropy function $s(F_{ij}, e, \psi_{ij}, \partial_k \psi_{ij}, \beta_{ij}, \partial_k \beta_{ij})$ is given.

\subsection{{Boundary conditions}}
It is remarkable that one can get natural boundary conditions for  internal variables with the requirement that the {intrinsic} part of  entropy {flux} \re{entc} is zero at the boundary. There are three basic possibilities:
\begin{enumerate}
 \item \emph{Combined condition}. In this case $\partial_{\partial_i\psi_{lm}}s$ and $\partial_{\partial_i\beta_{lm}}s$  are orthogonal to  $f_{lm} $ and $g_{lm}$.
 \item \emph{No change condition}. Internal variables do not change at the boundary, $\dot \psi_{ij}$ and $\dot \beta_{ij}$ are zero.
 \item \emph{Gradient condition}. $t_i \partial_{\partial_i\psi_{lm}}s=0_{lm}$ and $t_i \partial_{\partial_i\beta_{lm}}s=0_{lm}$ for any vector $t_i$ that is parallel to the boundary. In the case of a quadratic dependence on gradients, the condition is that  gradients of internal variables are orthogonal to the boundary.
\end{enumerate}

\subsection{Remarks on reciprocity relations}

Classical irreversible thermodynamics requires special conductivity matrices {obtained} by reciprocity relations.
In the case of symmetric matrices, the reciprocity relations are of Onsager type \cite{Ons31a1,Ons31a2}, if the matrices are antisymmetric, then they are {of} Casimir type \cite{Cas45a}.
These restrictions are justified by arguments from statistical physics requiring a microscopic interpretation of  thermodynamic variables.
If these variables are even or odd functions of the microscopic velocities, then the conductivity matrix is symmetric or antisymmetric, respectively.
However, in our case one cannot specify the conductivity matrices, the conditions of Onsager or Casimir type reciprocity relations are not fulfilled.
In the following we will show some possible interpretations of the dual internal variables, and we will see that the most straightforward interpretation, {namely,} the micromorphic generalized mechanics when the internal variables are of deformation and deformation rate type, requires general forms of the
conductivity matrices.
Moreover, the background of internal variables in continuum mechanics is originated in structural changes in a material, but not on the microscopic, atomic, {or} mesoscopic level \cite{VanAta08a}.
Therefore we simply do not see any reasons to further specialize the theory and we keep our treatment universal, independent of microscopic or mesoscopic interpretations.


Ideal{ized} theories are characterized by a nondissipative {behavior}.
In the case of heat conduction, the heat conduction coefficient $\lambda$ is zero
and, therefore, the heat flux, the {flux}  of the internal energy density, is zero, too.
In the case of pure mechanical interaction without internal variables, the ideal{ized} theory is characterized by zero viscosities, the bulk viscosity $l^{11}=\eta_v=0$ and the shear viscosity $s^{11}=2\eta=0$ are zero, together with the condition $a^{11}=0$.
Therefore, the viscous stress is zero, the stress in the momentum balance is the static stress $ t^{ij} =- \varrho T  \partial_{F_{ij}}s$.
However, antisymmetric {terms of} conductivity {matrices} do not produce entropy, therefore in our case the coupling {between} different terms may result in nondissipative transport, too.

\section{Spatial representation}

In {this section, we represent governing} equations {in spatial} framework and at the same time introduce {the} small {strain} approximation.
Technical details are given in the Appendix.

We can transform balances of internal energy and momentum \re{mom_bal}, \re{inte_bal} into {spatial} form due to the Nanson theorem,
to obtain
\eqn{linte_bal}{
\varrho \dot{e} + \partial_i  q_i &=  t_{ij} \partial_jv_i, \nl{lmom_bal}
\varrho \dot{v}_i - \partial_j  t_{ij} &= 0.
}
Here $\varrho\approx \varrho_0$ is the density {in the actual configuration}, $t_{ij}$ is the Cauchy stress {tensor}, and $q_i$ is the {spatial flux} of the internal energy, the upper dot denotes the substantial time derivative, $\dot e = \partial_t e +v^i\partial^i e$.
{Spatial} and material forms of these balances are identical 
in {the small strain approximation}. 

The {spatial form of} evolution equations of  internal variables  is the following:
\eqn{levol}{
  \stackrel{\lozenge}{\psi}_{ij}  + f_{ij}=0, \qquad
  \stackrel{\lozenge}{\beta}_{ij} + g_{ij}=0.
}
{The symbol} $\stackrel{\lozenge}{}$ {denotes} the spatial form of the material time derivative of a second-order tensor.
 { The material time derivative is an upper convected one, because internal variables are defined on the material manifold, supposedly independent of the motion of the continuum (see Appendix):}
\eqn{PKtder}{
\stackrel{\lozenge}{\psi}_{ij} = \dot \psi_{ij} -\partial_k v_i \psi_{kj} - \partial_kv_j\psi_{jk},
}
{i.e., spatial and material forms of constitutive functions for internal variables are the same in the small strain approximation if velocity gradients are negligible}.
The {spatial} form of the entropy {flux} is {represented as}
\eqn{lentc}{
J_i = \partial_{e}s q_{i} +
  \varrho \partial_{\partial_i\psi_{lm}}s\   f_{lm} +
  \varrho \partial_{\partial_i\beta_{lm}}s\  g_{lm} + J_i^0.
}
We do not deal {here} with any interpretation of the extra entropy flux $J_i^0$ and, therefore, we assume that it is {equal to} zero.
This {requirement} is not necessary, {and} there are phenomena {which} can be modelled only with the help of nonzero extra entropy flux \cite{CimVan05a,CiaAta07a,VanFul12a}.

Finally, the {spatial} form of the entropy production in the small strain approximation is {the following}:
\eqn{lentrpr}{
\begin{split}
   & \sigma=\partial_i \left( \frac{1}{T} \right)  q_i +
\frac{1}{T}\left(  t_{ij} +
  \varrho T  \partial_{\epsilon_{ij}}s \right)\dot\epsilon_{ij} -  
  \\
    &  -  \left(\partial_{\psi_{ij}}s - \partial_k \partial_{\partial_k\psi_{ij}}s  \right)\varrho f_{ij} - \left(\partial_{\beta_{ij}}s - \partial_k \partial_{\partial_k\beta_{ij}}s \right)\varrho g_{ij} \geq 0.
\end{split}
}
Here $\epsilon_{ij}$ is the Cauchy {strain}. 

The original  dependency of the entropy function {on the deformation gradient} $F_{ij}$ turns to {a} dependency {on} $\epsilon_{ij}$ because of the small strain approximation.
Moreover, it is convenient to introduce the specific free energy function instead of the specific entropy as {a} thermodynamic potential.
In our case, the {expression} of {the corresponding} partial Legendre transformation is:
\eqn{freee}{
w(T,\epsilon_{ij},\psi_{ij},\partial_k\psi_{ij},\beta_{ij},\partial_k\beta_{ij}) =
e - Ts(e,\epsilon_{ij},\psi_{ij},\partial_k\psi_{ij},\beta_{ij},\partial_k\beta_{ij}),
}
Therefore,
 \eqn{der}
 {
 \f{\partial s}{\partial e} = \f{1}{T}\quad \mbox{and} \quad \f{\partial w}{\partial T} = -s,
 }
 keeping fixed {all} other variables.
 Other partial derivatives of $w$ and $s$ are related as follows:
\eqn{pdw}{
\left. \f{\partial s}{\partial \epsilon_{ij}}\right|_e = \left.-\f{1}{T} \f{\partial w}{\partial \epsilon_{ij}}\right|_T, \quad
\left. \f{\partial s}{\partial \psi_{ij}}\right|_e = \left. -\f{1}{T} \f{\partial w}{\partial \psi_{ij}}\right|_T, \quad
\left. \f{\partial s}{\partial \beta_{ij}}\right|_e = \left.-\f{1}{T} \f{\partial w}{\partial \beta_{ij}}\right|_T, \nnl{pdw1}
\left. \f{\partial s}{\partial \partial_k\psi_{ij}}\right|_e = \left.-\f{1}{T} \f{\partial w}{\partial \partial_k\psi_{ij}}\right|_T, \quad
\left. \f{\partial s}{\partial \partial_k\beta_{ij}}\right|_e = \left.-\f{1}{T} \f{\partial w}{\partial \partial_k\beta_{ij}}\right|_T.
}
{All} not indicated variables are kept fixed in the partial derivatives.
In terms of the free energy function, we can transform entropy production \re{lentrpr} to the  form
\eqn{ldiss}{
\begin{split}
   & T\sigma=\partial_i \ln T \left(q_i +
  \varrho f_{ij} \partial_{\partial_k\psi_{ij}}w +
    \varrho g_{ij} \partial_{\partial_k\beta_{ij}}w \right) +
    \left(t_{ij} - \varrho\partial_{\epsilon_{ij}}w \right)\dot \epsilon_{ij} + 
    \\
    & +\left(\partial_{\psi_{ij}}w -
    \partial_k(\partial_{\partial_k\psi_{ij}}w)\right)\varrho f_{ij} +
   \left(\partial_{\beta_{ij}}w -
    \partial_k(\partial_{\partial_k\beta_{ij}}w)\right)\varrho g_{ij}  \geq 0.
\end{split}
}
It is worth to introduce thermodynamic forces and fluxes {in the small strain approximation} according to {Eq.} \re{ldiss}:
\begin{center}
{\scriptsize
\begin{tabular}{c|c|c|c|c}
       & Thermal & Mechanical & Internal 1 & Internal 2 \\ \hline
Fluxes & $\hat q_k =  q_k + $ &
    $\hat t^v_{ij} = $ &
    $\varrho \hat f_{ij}$ &
    $\varrho \hat g_{ij}$\\
       & $\varrho f_{ij} \partial_{\partial_k\psi_{ij}}w+
	\varrho g_{ij} \partial_{\partial_k\beta_{ij}}w$  &
       $t_{ij} - \varrho \partial_{\epsilon_{ij}} w$ & &\\ \hline
Forces &$\partial_k \ln T$ &
    $\dot \epsilon_{ij}$ &
    $\hat X_{ij} = $ &
    $\hat Y_{ij} = $\\
    & & & $\partial_{\psi_{ij}}w - \partial_k  \partial_{\partial_k\psi_{ij}}w $&
    $\partial_{\beta_{ij}}w - \partial_k  \partial_{\partial_k\beta_{ij}}w$
\end{tabular}\\
\vskip .21cm
}
{Table 2. Thermodynamic fluxes and forces in the small {strain approximation} 
}\end{center}
It is easy to see that 
thermodynamic forces and fluxes {in the small strain approximation are very similar to those in the PK frame}. 
The most important difference is the regrouping of  terms {which are} proportional to the temperature gradient.
This {representation of forces and fluxes} is more convenient {for the separation (or coupling) of} thermodynamic and mechanical parts of the {entropy production}, especially in the case of thermal stresses.
The {solution of dissipation inequality \re{ldiss} is provided by} conductivity equations:
\eqn{lOns1}{
\hat{q}_k &= q_k +\varrho f_{ij} \partial_{\partial_k\psi_{ij}}w +
  \varrho g_{ij} \partial_{\partial_k\beta_{ij}}w &=
  \lambda \partial_k \ln T, \nl{lOns2}
\hat t^v_{ij} &=  t_{ij} - \varrho\partial_{\epsilon_{ij}} w &=
  \hat L^{11}_{ijkl} \dot \epsilon_{kl} +
  \hat L^{12}_{ijkl} \hat X_{kl} +
  \hat L^{13}_{ijkl} \hat Y_{kl}, \nl{lOns3}
\varrho \stackrel{\lozenge}{\psi}_{ij} &= \varrho \hat f_{ij} &=
  \hat L^{21}_{ijkl} \dot \epsilon_{kl} +
  \hat L^{22}_{ijkl} \hat X_{kl} +
  \hat L^{23}_{ijkl} \hat Y_{kl}, \nl{lOns4}
\varrho \stackrel{\lozenge}{\beta}_{ij} &= \varrho \hat g_{ij} &=
  \hat L^{31}_{ijkl} \dot \epsilon_{kl} +
  \hat L^{32}_{ijkl} \hat X_{kl} +
  \hat L^{33}_{ijkl} \hat Y_{kl}.
}
In the case of isotropic materials, a decomposition of conductivity matrices can be performed  introducing symmetric, antisymmetric and spherical parts of corresponding tensorial forces and {fluxes similarly to previous case}:
\eqn{liso1}{
 \hat L^{IJ}_{ijkl} =
  s^{IJ} \delta_{i\langle k}\delta_{jl\rangle} +
  a^{IJ} \delta_{i[k}\delta_{jl]} +
  l^{IJ} \delta_{ij}\delta_{kl}.
}
In the following, we treat different special cases, all of them with constant conductivity matrices.
%
%

\section{{Examples}}

\subsection{Linear viscoelasticity, relaxation, and Ginzburg-Landau equation}
We {start with} the simplest case when there is not any coupling between  evolution equations {\re{lOns1}--\re{lOns4}}.
In this case, the {balance}  equation of the {linear} momentum and  {evolution equations for} internal variables are independent.
%
{This means that }
 the conductivity hypermatrix is diagonal $\hat L^{12} =\hat L^{21} =\hat L^{13}=\hat L^{31}=\hat L^{32}=\hat L^{23} = 0$, {and we assume that}  the heat conduction coefficient $\lambda$ is zero.
The free energy is additively decomposed into parts  {which are dependent on $\psi_{ij}$, $\beta_{ij}$, and $\epsilon_{ij}$ separately.}
 Dissipation inequality \re{ldiss} reduces to
\eqn{ldissa}{
\begin{split}
   & T\sigma=
    \left(t_{ij} - \varrho\partial_{\epsilon_{ij}}w \right)\dot \epsilon_{ij} + \\
    &  + \left(\partial_{\psi_{ij}}w -
    \partial_k(\partial_{\partial_k\psi_{ij}}w)\right)\varrho f_{ij} +
   \left(\partial_{\beta_{ij}}w -
    \partial_k(\partial_{\partial_k\beta_{ij}}w)\right)\varrho g_{ij}  \geq 0,
\end{split}
}
and we see that terms related to internal variables are completely similar.
{This means that in the absence of couplings it is enough to analyze only one of them.}

For the viscous part of the stress {we have, therefore,}
\eqn{vs_linvis}{
t_{ij} -t^{ela}_{ij} = t_{ij}^v = l^{11} \dot \epsilon_{kk} \delta_{ij} + s^{11} \dot \epsilon_{(ij)}.
}
{where} the elastic stress {is introduced as usually}
\eqn{ess_linvis}{
\varrho \f{\partial w}{\partial \epsilon_{ij}} =  t^{ela}_{ij} = \varrho \lambda \epsilon_{kk} \delta_{ij} + \varrho 2\mu \epsilon_{ij},
}
$l^{11} = \eta_v$ corresponds to the bulk viscosity and $s^{11} = \eta$ is the shear viscosity.
{
All other coefficients in conductivity matrices are zero due to the absence of any coupling.}
The  evolution equation for the internal variable ${\psi}_{ij}$ results in:
\eqn{psiGL}{
\varrho \stackrel{\lozenge}{\psi}_{ij}  &= \hat L^{22}_{ijlm} \left(\partial_{\psi_{lm}}w - \partial_k  \partial_{\partial_k\psi_{lm}}w \right). 
}
{Assuming the isotropy of} the conductivity tensor  $\hat L^{22}$, {we can}  decompose Eq. (\ref{psiGL}) into six independent evolution equations for the spherical, symmetric traceless and antisymmetric parts of the internal variable {tensor}.
 These evolution equations for the internal variable $\psi_{ij}$  {give the} generalization of the Ginzburg-Landau-Khalatnikov equation, introduced first  {as a scalar equation in the case of superconductors \cite{LanGin50a,LanKha54a}}.
 {For each} free energy functional,
  the structure of  {such} equations is universal and widely used  {with} different thermodynamic arguments \cite{Grm93a,GrmOtt97a,Fab99a,FabAta03a}.
  The Ginzburg-Landau equation was derived by pure thermodynamic arguments as the evolution equation for a second-order weakly nonlocal internal variable in \cite{Van05a}
   {(see also \cite{BerEta11a})}.
    That derivation shows a universal character  {of the Ginzburg-Landau equation}: the second law requires an evolution equation of this form for an internal
variable without any other constraints independently of the microscopic background.

The symmetric traceless form  {part of Eq. \re{psiGL}} gives the de Gennes-Landau theory of liquid crystals,  {if} a suitable quadratic dependence of the free energy {on}  $\psi_{\langle ij\rangle}$ is introduced \cite{deGPro93b}.
This very particular example of  {the proposed approach shows the} rich{ness of the} mathematical structure and its physical  {interpretation}. 

\subsection{Generalized standard linear solid}

In this section we consider a coupling between a single tensorial internal variable and deformation, which results in the thermodynamic theory of rheology \cite{Ful12m,Ver97b}.  Therefore, we {still} assume zero heat conduction coefficient and  $\hat L^{13}=\hat L^{31}=\hat L^{32}=\hat L^{23} = 0$.
{However}, {$\hat L^{12} \neq 0$ and $ \hat L^{21} \neq 0$ }anymore.
Let us further reduce the treatment by introducing a local theory for the single internal variable with the simplest quadratic dependence of the free energy on the internal variable without its gradients.
In this case, the free energy can be written as
\eqn{wfun_PT}{
\begin{split}
   & w(\epsilon_{ij},\psi_{ij}) =
  \f{\lambda}{2} \epsilon_{ii}^2 +  \mu \epsilon_{ij}\epsilon_{ij} +  \\
    & +\f{b_1}{2} \psi_{ii}^2 +  \f{b_2}{2} \psi_{ij}\psi_{ij} + \f{b_3}{2} \psi_{ij}\psi_{ji} +
   g_{1} \psi_{ii}\epsilon_{jj} + g_{2} (\psi_{ij}+\psi_{ji})\epsilon_{ij}.
\end{split}
  }
Thermodynamic stability requires the convexity of the free energy, hence  the inequalities follow:
\eqn{wpos}{
\begin{split}
   & 3\mu+2\lambda\geq 0, \quad
\mu\geq 0, \quad
b_2\geq 0, \quad
b_2+b_3\geq 0, \quad
3b_1+2b_2 \geq 0,  \\
    & \mu(b_2+b_3) - g_{1}^2 \geq 0, \quad (3\lambda + 2\mu)(3 b_1 + b_2 + b_3)- (3 g_{1}+2 g_{2})^2 \geq 0.
\end{split}
}
The entropy production contains two tensorial terms that may be coupled:
\eqn{ldis}{
T\sigma=     \left(t_{ij} - \varrho\partial_{\epsilon_{ij}} w \right)\dot \epsilon_{ij}+
   \partial_{\psi_{ij}}w\varrho f^{ij}  \geq 0,
}
In isotropic materials, spherical, deviatoric and antisymmetric parts of the tensors are independent.
For the deviatoric, symmetric traceless part we obtain the following conductivity equations from Eqs. \re{wfun_PT} and \re{ldis}:
\eqn{PTcon1}{
t_{\langle ij \rangle} - 2\varrho(\mu \epsilon_{\langle ij \rangle} + g_{2} \psi_{\langle ij \rangle}) &=
 s^1 \dot \epsilon_{\langle ij \rangle} -
 s^{12}\left((b_2+b_3)\psi_{\langle ij \rangle} + 2 g_2 \epsilon_{\langle ij \rangle} \right) \nl{PTcon2}
\varrho\stackrel{\lozenge}{\psi}_{\langle ij \rangle} &=
 s^{21} \dot \epsilon _{\langle ij \rangle} -
 s^{2} \left((b_2+b_3)\psi_{\langle ij \rangle} + 2 g_2 \epsilon_{\langle ij \rangle} \right).
}
{Equation \re{PTcon1} is a constitutive equation for the deviatoric part of the stress $t_{\langle ij \rangle}$, and Eq. \re{PTcon2} is the evolution equation  of $\psi_{\langle ij \rangle}$.}
Moreover, the second law requires that the symmetric part of the conductivity matrix is positive definite, therefore,
\eqn{PTpos}{
s^1 \geq 0, \quad
s^2 \geq 0, \quad
s^1 s^2 - s^{12} s^{21} - \left(\f{s^{12}-s^{21}}{2}\right)^2 \geq 0.
}
The role of the internal variable may be better understood if we eliminate it from Eqs. \re{PTcon1}--\re{PTcon2}.
{Taking the material time derivative of Eq. \re{PTcon1} and substituting
$\stackrel{\lozenge}{\psi}_{\langle ij \rangle}$ from Eq. \re{PTcon2} and $\psi_{\langle ij \rangle}$ from Eq. \re{PTcon1} into the obtained form, the following relation follows:}

\eqn{PT}{
\stackrel{\lozenge}{t}_{\langle ij \rangle} & + s^2\rho (b_2+b_3)t_{\langle ij \rangle} = \nnl{PTt1}
&s^1 \ddot \epsilon_{\langle ij \rangle} +
\rho\left[(b_2+b_3)(2\mu  + (s^1s^2-s^{12}s^{21})+(s^{21}-s^{12})2 g_{2})\right]
  \dot\epsilon_{\langle ij \rangle} + \nnl{PTt2}
 &2  s^2\rho^2((b_2+b_3) \mu - 2 g^2) \epsilon_{\langle ij \rangle}
}
The positive definiteness of the free energy $w$ {requires that} the coefficient $(b_2+b_3) \mu - 2 g^2$ is non-negative and Eq.\re{PTt2}  can be transformed to
\eqn{PTs}{
\tau \stackrel{\lozenge}{t}_{\langle ij \rangle} + t_{\langle ij \rangle} =
\tau_d \ddot{\epsilon}_{\langle ij \rangle} +
2 \nu\dot{\epsilon}_{\langle ij \rangle} +
E \epsilon_{\langle ij \rangle},
}
where
\eqn{PTcoeff}{
\tau &= \frac{1}{\rho s^2 (b_2+b_3)}, \qquad
\tau_d = s^1 \tau,\quad
E = s^2\rho^2((b_2+b_3) \mu - 2 g^2)\tau,\nnl{PTcoefft}
\nu &= \rho\left[(b_2+b_3)(2\mu  + (s^1s^2-s^{12}s^{21})+(s^{21}-s^{12})2 g_{2})\right]\tau.
}
Constitutive relation \re{PTs} reduces to the Zener body of rheology or standard linear solid \cite{Fer80b,Tsc89b},
if $\tau_d= 0$ and  the nonlinear part of the time derivative is neglected.
This rheological model is widely used in different {fields} from biology \cite{KlaEta07a} to engineering \cite{HaaSlu01a}.
The complete form {of this constitutive relation (with  $\tau_d \neq 0$)} is called the \textit{inertial Poynting-Thomson body}.
The proposed thermodynamic model has  remarkable properties which are distinctive {in comparison with} more intuitive approaches:
\begin{enumerate}
 \item Neither the inertial term beyond the standard model, nor the coupled volumetric-deviatoric effect is neglectable in general, both are important, e.g. in experimental rock rheology \cite{MatTak93a,Mat08a,LinEta10p}, where an inertial Poynting-Thomson body is used for both deviatoric and spherical parts of the deformation in order to get a good agreement with  experimental data.
 \item Reciprocity {has not been required in the proposed approach. The same is true} in the simplest case of a standard linear solid body, otherwise the second law of thermodynamics contradicts to  observations \cite{Ful12m}.
 \item { The complete form of constitutive equation \re{PTcon1}-\re{PTcon2} is preferable instead of \re{PTs} if dynamical problems with the coupled balance of momentum \re{mom_bal} are needed to be solved.} This is an important advantage of the thermodynamic approach.
\end{enumerate}

\subsection{Dual internal variables}

Now we consider dual (coupled) tensorial internal variables which are independent of mechanical and thermal interactions  ($\lambda =0$, $\hat L^{11} =\hat  L^{12}= \hat L^{13}=  \hat L^{21}= \hat L^{31}=0$).
The evolution equations of the coupled tensorial internal variables {again follow from the dissipation inequality}.
In the small {strain approximation} and {  with small velocity gradients}, Eqs. \re{lOns3}-\re{lOns4} are simplified to
\eqn{dual1}{
\dot{\psi}_{ij} &= \hat L^{22}_{ijkl} X_{kl} + \hat L^{23}_{ijkl} Y_{kl}, \nl{dual2}
\dot{\beta}_{ij} &= \hat L^{32}_{ijkl} X_{kl} + \hat L^{33}_{ijkl} Y_{kl}.
}
Here $X_{ij} = \partial_{\psi_{ij}} w - \partial_k \left(\partial_{\partial_k\psi_{ij}}w\right)$, and
$Y_{ij} = \partial_{\beta_{ij}} w - \partial_k \left(\partial_{\partial_k\beta_{ij}}w\right)$.
Equations \re{dual1}--\re{dual2} are independent of balances of linear momentum and  energy, if  the free energy can be {decomposed} into a sum of functions {depending on}  two sets of variables, $e,\epsilon_{ij}$ and $\psi_{ij}, \beta_{ij}$ separately{, and if the objective time derivatives can be substituted by the substantial time derivative, that is the nonlinear terms are neglectable in the upper convected derivatives.}

In the case of isotropic materials,  tensorial equations \re{dual1}--\re{dual2} are decomposed into a spherical, deviatoric and antisymmetric parts with scalar coefficients.
The remarkable difference between the {evolution described by a}  Ginzburg-Landau-type equation based {on a single internal variable} and this dual structure becomes apparent after the separation of the symmetric and antisymmetric parts of the particular decompositions.

As an example, we consider completely decoupled deviatoric {evolution} equations.
Then  conductivity matrices are two-dimensional and  evolution equations are the following:
\eqn{ddual1}{
\dot \psi_{\langle ij\rangle} &= s_1 X_{\langle ij\rangle} + (s+a) Y_{\langle ij\rangle}, \nl{ddual2}
\dot \beta_{\langle ij\rangle} &= (s-a) X_{\langle ij\rangle} + s_2 Y_{\langle ij\rangle}.
}
The above {evolution} equations are decomposed into a symmetric part, which represents a dissipative evolution and therefore produces entropy, and  the antisymmetric part that does not produce entropy and represents a non-dissipative part of the evolution.
The role of the non-dissipative part {can be better} understood with the help of the following free energy function, where we assume a local theory for  $\beta_{\langle ij\rangle}$:
\eqn{wfun_dual}{
w(\beta_{\langle ij\rangle },\psi_{\langle ij\rangle},\partial_k\psi_{\langle ij\rangle}) =
  \f{\hat c}{2} \beta_{\langle ij\rangle}\beta_{\langle ij\rangle} +
    w_h(\psi_{\langle ij\rangle}) +
    w_g(\partial_k\psi_{\langle ij\rangle}) = \nnl{wfun_dual2}
  \f{\hat c}{2} \beta_{\langle ij\rangle}\beta_{\langle ij\rangle} +\f{\hat b}{2} \psi_{\langle ij\rangle}\psi_{\langle ij\rangle} +
    \f{f_1}{2} \partial_k\psi_{\langle ij\rangle}\partial_k\psi_{\langle ij\rangle} +
  \f{f_2}{2} \partial_k\psi_{\langle ij\rangle}\partial_i\psi_{\langle jk\rangle} +
  \f{f_3}{2} \partial_k\psi_{\langle ik\rangle}\partial_j\psi_{\langle ij\rangle} .
}
Here $w_h$ is the homogeneous, local {part} and $w_g$ is the gradient dependent, weakly nonlocal part of the free energy {related to} the variable $\psi_{\langle ij\rangle}$.
The {second line of Eq. \re{wfun_dual2}} shows a particular quadratic form of corresponding functions.
 If $f_2=f_3=0$, then we obtain the usual {second} gradient theory.
 The non-dissipative part of  evolution equations \re{ddual1}-\re{ddual2} {has the form}
\eqn{ddual1s}{
\dot \psi_{\langle ij\rangle} &= a \hat c \beta_{\langle ij\rangle}, \nl{ddual2s}
\dot \beta_{\langle ij\rangle} &= -a \left(\partial_{\psi_{\langle ij\rangle}} w_h -
\partial_k(\partial_{\partial_k\psi_{\langle ij\rangle}} w_g)\right)
}
It is easy to eliminate $\beta_{\langle ik\rangle}$ and obtain an evolution equation for $\psi_{\langle ik\rangle}$, {which} is second order in time:
\eqn{ELag}{
\frac{1}{a^2 \hat c} \ddot \psi_{\langle ik\rangle} +
  \partial_{\psi_{\langle ik\rangle}}w_h -
  \partial_k \left(\partial_{\partial_k\psi_{\langle ik\rangle}} w_g \right) = 0
}
The {last} equation can be considered as the Euler-Lagrange equation of the Lagrangian
\eqn{Lag}{
L(\psi_{\langle ik\rangle},\dot \psi_{\langle ik\rangle}) =
\f{1}{2a^2 \hat c} \dot \psi_{\langle ik\rangle}\dot \psi_{\langle ik\rangle} -
w(\psi_{\langle ik\rangle},\partial_k\psi_{\langle ik\rangle}).
}
It is remarkable that  natural thermodynamic boundary conditions of the zero entropy {flux} requirement correspond exactly to  natural boundary conditions of the variational principle.

\subsection{Dissipative generalized continua}

It was already observed by Berezovski, Engelbrecht and Maugin \cite{BerEta11a} that generalized thermomechanics {of solids} is a
{particular case} of {the} dual internal variables {theory}.
They observed that  evolution equations of the non-dissipative theory of Mindlin correspond exactly to non-dissipative evolution equations of {the} dual internal variables {theory}.
However, they did not perform a complete thermodynamic analysis and, therefore, their observation is restricted to the idealized non-dissipative case.

{The corresponding thermodynamic} analysis was performed in the second section of the present paper exploiting the dissipation inequality.
The coupled {evolution} equations {are represented in the form of}  linear conductivity equations.
If one of the internal variables is {interpreted} as a microdeformation, then our calculations are to be considered as a pure thermodynamic derivation of a generalized dissipative {continuum theory}.
The particular example of the theory of Mindlin arises then {under} the following conditions:
\begin{enumerate}
 \item No {thermal and viscous dissipation}.
 \item Pure antisymmetric coupling between  internal variables.
 \item Quadratic free energy {function}.
\end{enumerate}

Another generalized continuum theory has been introduced by  Eringen and Suhubi \cite{EriSuh64a}.
To compare the entropy production, let us consider for simplicity a continuum without additional internal variables in the small strain approximation and uniform temperature field. Then the entropy production in the Eringen-Suhubi theory \re{ep1} lack the first and the last terms and can be written with our notation as
\eqn{ldissam}{
T\sigma=
    \left(t_{ij} - \varrho\partial_{\epsilon_{ij}}w \right)\dot \epsilon_{ij} +
  \left(s_{ij}-\tau_{ij}-\varrho \partial_{\psi_{ij}}w \right)\dot\psi_{ij} +
   \left(\mu_{ijk} -\partial_{\partial_k\psi_{ij}}w\right)\partial_k\dot\psi_{ij} \geq 0.
}
Here the internal variable $\psi_{ij}$ is identified with the microdeformation gradient $\chi'_{ij}$ of the Eringen-Suhubi theory and therefore the material time derivative is the substantial time derivative due to the deformation interpretation.
For the comparison, let us repeat here our entropy production \re{ldissa}:
\eqn{ldissao}{
\begin{split}
   & T\sigma=
    \left(t_{ij} - \varrho\partial_{\epsilon_{ij}}w \right)\dot \epsilon_{ij} +     \\
    & -  \left(\partial_{\psi_{ij}}w -
    \partial_k(\partial_{\partial_k\psi_{ij}}w)\right)\varrho \stackrel{\lozenge}{\psi}_{ij} -
   \left(\partial_{\beta_{ij}}w -
    \partial_k(\partial_{\partial_k\beta_{ij}}w)\right)\varrho \stackrel{\lozenge}{\beta}_{ij}
    \geq 0.
\end{split}
}
{ 
The difference in the entropy flux in the two theories and the different concept of constitutive quantities determine the diversity in the entropy production.
We have introduced the evolution equations of internal variables as constitutive relations to be determined from the entropy inequality.
Eringen and Suhubi \cite{EriSuh64a} simply indicated the form of the dissipation inequality following from their definition of stresses.
}

Regarding the exploitation of the second law one should observe the following:
\begin{itemize}
 {\item The micromomentum balance and the evolution equation of $\psi_{ij}$ are not constructed from the dissipation inequality.}
 \item The micromomentum balance is not {used as} a constraint for the entropy inequality in the Eringen-Suhubi derivation. However, it is implicitly considered during the application of the Coleman-Noll procedure assuming that the multipliers of the time derivatives should be zero.
 \item The entropy {flux} is not an arbitrary constitutive function, but it is restricted to the classical $J_i = q_i/T$ in the case of Eringen and Suhubi.
 \item The constitutive state space is not weakly nonlocal and not fixed in the Eringen-Suhubi derivation.
\end{itemize}

As we have shown above, not only the basic assumptions, but also the final equations of the Eringen-Suhubi theory are particular and can be obtained from our generalized approach if several dissipative terms are neglected.
Virtual power approaches also introduce an entropy production that have a similar form and similar limitations as the Eringen-Suhubi approach has (see e.g. \cite{ForSie06a}).

\subsection{Heat conduction and weakly nonlocal internal variables: Microtemperature}

Finally, let us {consider the case with a} non-zero heat conduction coefficient $\lambda$.
{Neglecting the viscosity influence ($\hat L^{11} =\hat  L^{12}= \hat L^{13}=  \hat L^{21}= \hat L^{31}=0$),
we chose the free energy dependence on internal variables in the form of Eq. \re{wfun_dual2}, but with the reduced conductivity matrix for deviatoric evolution equations
\eqn{ddual1a}{
\dot \psi_{\langle ij\rangle} &=  a Y_{\langle ij\rangle}, \nl{ddual2a}
\dot \beta_{\langle ij\rangle} &= -a X_{\langle ij\rangle} + s_2 Y_{\langle ij\rangle}.
}
which corresponds to the choice $s=s_{1}=0$ in Eqs. \re{ddual1}-\re{ddual2}.
The free energy dependence \re{wfun_dual2} allows to represent the evolution equations in the form
\eqn{ddual1sa}{
\dot \psi_{\langle ij\rangle} &= a \hat c \beta_{\langle ij\rangle}, \nl{ddual2sa}
\dot \beta_{\langle ij\rangle} &= -a \left(\partial_{\psi_{\langle ij\rangle}} w_h -
\partial_k(\partial_{\partial_k\psi_{\langle ij\rangle}} w_g)\right)+s_{2}\hat c \beta_{\langle ij\rangle},
}
which can be reduced to the single second-order evolution equation for the primary internal variable $\psi_{\langle ij\rangle}$
\eqn{ELaga}{
\frac{1}{a^2 \hat{c}} \ddot \psi_{\langle ik\rangle} - \frac{s_{2}}{a}\dot \psi_{\langle ik\rangle}+
  \partial_{\psi_{\langle ik\rangle}}w_h -
  \partial_k \left(\partial_{\partial_k\psi_{\langle ik\rangle}} w_g \right) = 0,
}
{which is similar to the Jeffreys type modification of the Maxwell-Cattaneo-Vernotte equation \cite{JosPre89a}.}

The thermal part of the dissipation inequality is satisfied by the modified Fourier law that follows from Eq. \re{lOns1}
\eqn{lOns1a}{
q_k +\varrho f_{ij} \partial_{\partial_k\psi_{ij}}w +
  \varrho g_{ij} \partial_{\partial_k\beta_{ij}}w =
  \lambda \partial_k \ln T.
}
As it was shown  \cite{BerEta11a} on the example of one-dimensional thermoelasticity, the primary internal variable   $\psi_{\langle ij\rangle}$ can be interpreted in this case as a microtemperature.
In this context, it is understood as a fluctuation of the macrotemperature due to the influence of the existing microstructure.
}
 The solution of the equations shows that influence of microtemperature may result in a wavelike propagation of {temperature} if the corresponding damping effects are small \cite{BerEta11a}.

\section{Summary and discussion}

The paper is devoted to  the answer of the following question:
How could we obtain evolution equations of physical quantities, {about which} we do not know anything, {i.e.,} only general principles can be considered? There are essentially two basic approaches.
The first one postulates a variational principle of Hamiltonian type coming from mechanics.
In this case dissipation is something additional to the non-dissipative basic mechanical evolution.
The second approach is coming from thermodynamics: one can assume that the evolution of  new variables is not exception from the second law and generate their governing differential equations accordingly.
This is the situation in the case of internal variables in general as it was summarized by Maugin and Muschik \cite{MauMus94a1,MauMus94a2}.
The two approaches can be {generalized}.
Thermodynamic principles and dual internal variables in the framework of a second order weakly nonlocal theory give a straightforward and simple way of {the generalization} \cite{VanAta08a}.

{As a result,}
the thermodynamic consistency of continuum mechanics with dual tensorial internal variables was analyzed {in the present paper by} the  Liu {approach to the exploitation of the second law} in the Piola-Kirchhoff framework.
Then local evolution was considered in  isotropic materials in the small strain approximation.
The  entropy production was calculated and thermodynamic forces and fluxes were identified.
Then a quadratic free energy and linear  conductivity relations closed the system of equations.
The final evolution equations in a non-dissipative case are equivalent of those for micromorphic continua, therefore, the thermodynamic method gave a dissipative extension of the original Mindlin theory.

We have given several particular examples that arise as special cases of the general theory.
Our goal was only partially a justification, but also the identification of the most important differences from other theories and the interpretation of some qualitative predictions of our approach.
We have seen that  phenomena of microtemperature,  sophisticated couplings in generalized rheology, and special properties of the dissipative extension without assuming reciprocity relations, all are open for experimental testing.
We think that this approach is essential in the case of generalized continua, where additional  coefficients are considered hardly measurable.
  In this respect we have analyzed the limitations of the dissipation in other classical approaches, in particular in the Eringen-Suhubi theory, which is one of the most developed from this point of view. We have seen, that due to the restrictive starting assumptions (mechanical interpretation of the internal variable, locality, special entropy flux, etc.) the considered dissipation is extremely limited. For example, under the traditional approaches one cannot recover neither the Ginzburg-Landau equation, nor simple viscoelasticity.

It is remarkable, that the finite deformation part of our approach shares  shortcomings of the Piola-Kirchhoff framework.
There are  indications that the requirement of objectivity and material frame indifference are not treated properly in this case \cite{FulVan12a,NolSeg10a}.
Beyond the reservations of using material manifolds in general, it is also remarkable that the exclusion of velocity field $v_i$ from the constitutive state space is not necessary \cite{MatVan06a,VanPap10a}.

We expect several interesting phenomena by the analysis of higher-order nonlocality at the mechanical and thermal side.
Here the comparison to phase field approaches looks like a promising direction (see e.g. \cite{FabMor03a,AndEta98a,God11a}).

{
\section{Appendix}

In this section we shortly derive the material time derivatives, introduce the small strain approximation and describe the transformation of the balances between a Piola-Kirchhoff and local frameworks. Here we distinguish between contra and covariant {as well as between}  material manifold  and space-time vectors and tensor components. The covariant and contravariant vectors are denoted by lower and upper indices, the space vector and tensor components are denoted by minuscules, and the vector and tensor components at the material manifold by capital letters. We assume here that the reference configuration is relaxed, stress free, therefore the transformation between the material and spatial descriptions is standard (see e.g. \cite{GurEta10b}). For a more detailed kinematics, considering general bodies, see \cite{FulVan12a}.

The material vectors are denoted by $X^i$, the spatial ones by $x^i$.
Therefore, the deformation gradient, the material manifold derivative of the motion, $\chi^i(t,X^J)$, is given as:
\eqn{defgrad}{
F^i_{\ J} = \partial_J \chi^i.
}
The transformation between material and spatial vectors and covectors is the following:
\eqn{vctraf}{
a^J = (F^{-1})^J_{\ i}a^i, \qquad
a^i = F^i_{\ J}a^J, \nl{hehe}
b_J = F^i_{\ J}b_i, \qquad
b_i = (F^{-1})^J_{\ i}b_J
}
In particular, the transformation of space derivatives follows the lower indexed covector rule:
\eqn{dertraf}{
\partial_J = F^i_{\ J}\partial_i, \qquad
\partial_i = (F^{-1})^J_{\ i}\partial_J.
}
The summation over repeated indices is still assumed.

\subsection{Material time derivatives}
The transformation of time derivatives is different for quantities with different tensorial character. For scalars, the partial time derivative on the material manifold, $\tilde \partial_t$, is the substantial derivative for local quantities.
We use the convenient dot notation for local quantities that corresponds to the partial time derivative on the material manifold for scalars, for the velocity and for acceleration fields:
\eqn{timeder_trafo}{
\tilde \partial_t a(t,X^i) = \dot a = \partial_t a(t,x^i)  + v^i\partial_i a(t,x^i), \nl{velacc}
\tilde \partial_t \chi^i = v^i, \qquad
\tilde \partial_{tt} \chi^i = \tilde \partial_t v^i =\dot v^i.
}
The material time derivative of internal variables with various tensorial character differ from each other.
Here we give the calculation for the second-order tensorial variable,  $\psi^{ij}$.
According to the definition of material time derivatives, the partial time derivative on the material manifold expressed by spatial fields gives the material time derivative:
\eqn{intvbal_trfo}{
  \tilde \partial_t{\psi}^{IJ} = &\tilde \partial_t\left( (F^{-1})^I_{\ i}(F^{-1})^J_{\ j}\psi^{ij} \right) \nonumber \\= &
  (F^{-1})^I_{\ i}(F^{-1})^J_{\ j}\dot\psi^{ij} -
  (F^{-1})^I_{\ k}\partial_t F^k_{\ L}(F^{-1})^L_{\ i}(F^{-1})^J_{\ j}\dot\psi^{ij}  \nonumber\\ \qquad -&
  (F^{-1})^I_{\ i}(F^{-1})^J_{\ k}\partial_t F^k_{\ L}(F^{-1})^L_{\ j}\dot\psi^{ij} \nonumber \\ = &
  (F^{-1})^I_{\ i}(F^{-1})^J_{\ j}\left( \dot\psi^{ij} - \partial_kv^i \psi^{kj}-\partial_kv^j \psi^{ik} \right).
}
Therefore, the spatial form of the abovementioned formula, the material time derivative of the tensor, is given as
\eqn{tenmatder}{
\stackrel{\lozenge}{{\psi}^{ij}}
=F^i_{\ I}F^j_{\ J}\tilde\partial_t\psi^{IJ} =
\dot\psi^{ij} - \partial_kv^i \psi^{kj}-\partial_kv^j \psi^{ik}.
}
Here we used the kinematic relation for the spatial velocity gradient and the time derivative of the deformation gradient $\partial_jv^i = \partial_tF^i_{\ J} (F^{-1})^J_{\ j}$. In case of cotensors or mixed tensors, the spatial form of the material time derivative is different.

\subsection{Small strains}
The small spatial strains are defined with the left Cauchy-Green deformation $A^{ij} = F^i_{\ J}F_J^{\ j}$, as
\eqn{sd}{
\epsilon^{ij} := \frac{1}{2}(A^{ij}-\delta^{ij}).
}
This choice is the best considering the requirement of objectivity \cite{FulVan12a}.
The material { time derivative of the strain in the} small strain approximation is the symmetric part of the velocity gradient
\eqn{eptder}{
\dot\epsilon^{ij} =\frac{1}{2}\dot A^{ij} \approx \frac{1}{2}(\partial^iv^j+\partial^jv^i).
}
It should be noted that in the small strain approximation one may obtain identical results starting from different deformation concepts.


The spatial form of the second material time derivative of the strain is particular
\eqn{matder_ep}{
\ddot \epsilon^{ij} = \frac{1}{2} \ddot A^{ij} \approx \partial^{(l}\dot v^{i)} + \partial_kv^i\partial_kv^j.
}
Here we have used that $\partial_l\dot v^i A^{lj} = \ddot F^i_J (F^{-1})^J_{\ l} F^l_{\ K}F_k^{\ j}.$

\subsection{Spatial balances}

The relation {between the} local density $\varrho$ and the material density $\varrho_0$ is
$$
\varrho = \f{\varrho_0}{\det F},
$$
{where $\det F$ is the determinant of $F^i_{\ K}$}

The local and Piola-Kirchhoff forms of the heat {flux} and the stress are, {respectively,}
\eqn{heat_trafo}{
q^i & = (\det F)^{-1} F^i_{\ J} q^J, \qquad
q^I  = \det F (F^{-1})^{\ I}_j q^j, \nl{stress_trafo}
t^{ij} & = (\det F)^{-1} F^i_{\ K} t^{Kj}, \qquad
t^{Ij}  = \det F (F^{-1})^I_{\ k} t^{kj}.
}
For the transformation of basic balances, the Nanson theorem is essential. {It} can be written as:
\eqn{Nanson}{
\partial_J\left(\det F (F^{-1})^J_{\ i}\right) = 0_i.
}
The proof is straightforward, when considering that the derivative of the determinant is $\partial_J (\det F) = \det F (F^{-1})^K_{\ l}\partial_J F^l_{\ K}$ and the derivative of the inverse deformation gradient is $\partial_J (F^{-1})^I_{\ j}= -(F^{-1})^I_{\ l}\partial_J F^l_{\ K}(F^{-1})^K_{\ j} $.

Then the transformation of the balance of internal energy follows by substituting the definitions:
\eqn{intebal_trafo}{
\begin{split}
  \varrho_0 \dot e + \partial_K q^K & =  \det F \varrho \dot e + \det F (F^{-1})^I_{\ j}\partial_I q^j =\\
    & =\det F(\varrho \dot e + \partial_jq^j) = t^{Ij} \partial_Iv^j = \det F\ t^{kj} \partial_k v^j.
\end{split}
}
Similarly, the balance of {linear} momentum {can be} easily obtained:
\eqn{mombal_trafo}{
\varrho_0 \dot v^i + \partial_K t^{Ki} = ... =
\det F (\varrho \dot v^i + \partial_k t^{ki} ) = 0^i.
}
Therefore, we obtain usual local balances of internal energy and momentum \re{linte_bal} and \re{lmom_bal} without approximations.}

\section{Acknowledgement}

The work was supported by the grants Otka K81161, K104260 and TT 10-1-2011-0061/ZA-15-2009. The authors thank Tam\'as F\"ul\"op and Csaba Asszonyi for valuable discussions.

\bibliographystyle{unsrt}

\begin{thebibliography}{10}

\bibitem{Min64a}
R.~D. Mindlin.
\newblock Micro-structure in linear elasticity.
\newblock {\em Archive for Rational Mechanics and Analysis}, 16:51--78, 1964.

\bibitem{EriSuh64a}
A.C. Eringen and E.S. Suhubi.
\newblock Nonlinear theory of simple micro-elastic solids {I.}
\newblock {\em International Journal of Engineering Science}, 2:189--203, 1964.

\bibitem{Ger73a}
P.~Germain.
\newblock The method of virtual power in continuum mechanics. {P}art 2:
  Microstructure.
\newblock {\em SIAM Journal of Applied Mathematics}, 25:556--575, 1973.

\bibitem{Eri92a}
A.~C. Eringen.
\newblock Balance laws of micromorphic continua revisited.
\newblock {\em International Journal of Engineering Science}, 30:805--810,
  1992.

\bibitem{Eri99b}
C.~Eringen.
\newblock {\em Microcontinuum Field Theories I. Foundations and Solids}.
\newblock Springer-Verlag, Berlin-etc.., 3th edition, 1999.

\bibitem{LeeWan11a}
J.~D. Lee and X.~Wang.
\newblock Generalized {M}icromorphic solids and fluids.
\newblock {\em International Journal of Engineering Science}, 49:1378--1387,
  2011.

\bibitem{ForEta08a}
S.~Forest and M.~Amestoy.
\newblock Hypertemperature in thermoelastic solids.
\newblock {\em Comptes Rendus Mecanique}, 336:347--353, 2008.

\bibitem{AslFor11c}
O.~Aslan and S.~Forest.
\newblock The micromorphic versus phase field approach to gradient plasticity
  and damage with application to cracking in metal single crystals.
\newblock In René de~Borst and Ekkehard Ramm, editors, {\em Multiscale Methods
  in Computational Mechanics}, Lecture Notes in Applied and Computational
  Mechanics, pages 135--154. Springer, 2011.

\bibitem{PapFor06a}
C.~Papenfuss and S.~Forest.
\newblock Thermodynamical frameworks for higher grade material theories with
  internal variables or additional degrees of freedom.
\newblock {\em Journal of Non-Equilibrium Thermodynamics}, 31(4):319--353,
  2006.

\bibitem{Hau93a}
P.~Haupt.
\newblock {\em Non-Equilibrium Thermodynamics with Applications to Solids},
  volume 336 of {\em Courses and Lectures}, chapter Thermodynamics of Solids.
\newblock Springer Verlag, Wien, New York, 1993.
\newblock CISM-Course, Udine 1992, p65 - 138.

\bibitem{Gya70b}
I.~Gyarmati.
\newblock {\em Non-equilibrium Thermodynamics /{F}ield Theory and Variational
  Principles/}.
\newblock Springer Verlag, Berlin, 1970.

\bibitem{VanMus94a}
P.~V\'an and W.~Muschik.
\newblock The structure of variational principles in nonequilibrium
  thermodynamics.
\newblock In J.~Verh\'as, editor, {\em Periodica Polytechnica, Physics and
  Nuclear Sciences}, volume 2/1-2, pages 111--122, 1994.
\newblock Minisymposium on Variational Methods in Thermophysics, 1994 in
  Berlin.

\bibitem{VanNyi99a}
P.~V\'an and B.~Ny\'\i{}ri.
\newblock Hamilton formalism and variational principle construction.
\newblock {\em Annalen der Physik (Leipzig)}, 8:331--354, 1999.

\bibitem{IrvKir50a}
J.~H. Irwing and J.~G. Kirkwood.
\newblock The statistical mechanical theory of transport processes. {IV}. the
  equations of hydrodynamics.
\newblock {\em J. Chem. Phys.}, 18:817--829, 1950.

\bibitem{BleMus91a}
S.~Blenk and W.~Muschik.
\newblock Orientational balances for nematic liquid crystals.
\newblock {\em Journal of Non-equilibrium Thermodynamics}, 16:67--87, 1991.

\bibitem{EhrAta97a}
H.~Ehrentraut, W.~Muschik, and C.~Papenfuss.
\newblock Mesoscopically derived orientation dynamics of liquid crystals.
\newblock {\em Journal of Non-Equilibrium Thermodynamics}, 22:285--298, 1997.

\bibitem{MauMus94a1}
G.~A. Maugin and W.~Muschik.
\newblock Thermodynamics with internal variables. {P}art {I}. {G}eneral
  concepts.
\newblock {\em Journal of Non-Equilibrium Thermodynamics}, 19:217--249, 1994.

\bibitem{MauMus94a2}
G.~A. Maugin and W.~Muschik.
\newblock Thermodynamics with internal variables. {P}art {II}. {A}pplications.
\newblock {\em Journal of Non-Equilibrium Thermodynamics}, 19:250--289, 1994.

\bibitem{ColGur67a}
B.~D. Coleman and M.~E. Gurtin.
\newblock Thermodynamics with internal state variables.
\newblock {\em The Journal of Chemical Physics}, 47(2):597--613, 1967.

\bibitem{MulWei12a}
I.~M\"uller and W.~Weiss.
\newblock Thermodynamics of irreversible processes - past and present.
\newblock {\em Eur. Phys. J. H}, 37:139--236, 2012.

\bibitem{VanAta08a}
P.~V\'an, A.~Berezovski, and J.~Engelbrecht.
\newblock Internal variables and dynamic degrees of freedom.
\newblock {\em Journal of Non-Equilibrium Thermodynamics}, 33(3):235--254,
  2008.
\newblock cond-mat/0612491.

\bibitem{Van04a}
P.~V\'an.
\newblock Weakly nonlocal continuum theories of granular media: restrictions
  from the {S}econd {L}aw.
\newblock {\em International Journal of Solids and Structures},
  41(21):5921--5927, 2004.
\newblock (cond-mat/0310520).

\bibitem{VanFul06a}
P.~V\'an and T.~F\"ul\"op.
\newblock Weakly nonlocal fluid mechanics - the {S}chr\"odinger equation.
\newblock {\em Proceedings of the Royal Society, London A}, 462(2066):541--557,
  2006.
\newblock (quant-ph/0304062).

\bibitem{Cim07a}
V.~A. Cimmelli.
\newblock An extension of {L}iu procedure in weakly nonlocal thermodynamics.
\newblock {\em Journal of Mathematical Physics}, 48:113510, 2007.

\bibitem{BerEta11a}
A.~Berezovski, J.~Engelbrecht, and G.~A. Maugin.
\newblock Generalized thermomechanics with dual internal variables.
\newblock {\em Archive of Applied Mechanics}, 81(2):229--240, 2011.

\bibitem{Wal84b}
R.~M. Wald.
\newblock {\em General Relativity}.
\newblock The University of Chicago Press, Chicago and London, 1984.

\bibitem{MatVan06a}
T.~Matolcsi and P.~V\'an.
\newblock Can material time derivative be objective?
\newblock {\em Physics Letters A}, 353:109--112, 2006.
\newblock math-ph/0510037.

\bibitem{Van05a}
P.~V\'an.
\newblock Exploiting the {S}econd {L}aw in weakly nonlocal continuum physics.
\newblock {\em Periodica Polytechnica, Ser. Mechanical Engineering},
  49(1):79--94, 2005.
\newblock (cond-mat/0210402/ver3).

\bibitem{Van09a1}
V\'an P.
\newblock Weakly nonlocal non-equilibrium thermodynamics - variational
  principles and {S}econd {L}aw.
\newblock In Ewald Quak and Tarmo Soomere, editors, {\em Applied Wave
  Mathematics (Selected Topics in Solids, Fluids, and Mathematical Methods)},
  chapter III, pages 153--186. Springer-Verlag, Berlin-Heidelberg, 2009.
\newblock (arXiv:0902.3261).

\bibitem{Ons31a1}
L.~Onsager.
\newblock Reciprocal relations of irreversible processes {I}.
\newblock {\em Physical Review}, 37:405--426, 1931.

\bibitem{Ons31a2}
L.~Onsager.
\newblock Reciprocal relations of irreversible processes {II}.
\newblock {\em Physical Review}, 38:2265--2279, 1931.

\bibitem{Cas45a}
H.~G.~B. Casimir.
\newblock On {O}nsager's principle of microscopic reversibility.
\newblock {\em Reviews of Modern Physics}, 17:343--350, 1945.

\bibitem{CimVan05a}
V.~A. Cimmelli and P.~V\'an.
\newblock The effects of nonlocality on the evolution of higher order fluxes in
  non-equilibrium thermodynamics.
\newblock {\em Journal of Mathematical Physics}, 46(11):112901--15, 2005.
\newblock cond-mat/0409254.

\bibitem{CiaAta07a}
V.~Ciancio, V.~A. Cimmelli, and P.~V\'an.
\newblock On the evolution of higher order fluxes in non-equilibrium
  thermodynamics.
\newblock {\em Mathematical and Computer Modelling}, 45:126--136, 2007.
\newblock cond-mat/0407530.

\bibitem{VanFul12a}
P.~V\'an and T.~F\"ul\"op.
\newblock Universality in heat conduction theory: weakly nonlocal
  thermodynamics.
\newblock {\em Annalen der Physik}, 524(8):470--478, 2012.
\newblock arXiv:1108.5589.

\bibitem{LanGin50a}
L.~D. Landau and V.~L. Ginzburg.
\newblock K teorii sverkhrovodimosti.
\newblock {\em Zhurnal Eksperimentalnoi i Teoreticheskoi Fiziki}, 20:1064,
  1950.
\newblock English translation: On the theory of superconductivity, in:
  Collected papers of L. D. Landau, ed. D. ter Haar, (Pergamon, Oxford, 1965),
  pp. 546-568.

\bibitem{LanKha54a}
L.~D. Landau and I.~M. Khalatnikov.
\newblock Ob anomal'nom pogloshchenii zvuka vblizi tochek fazovogo perekhoda
  vtorogo roda.
\newblock {\em Dokladu Akademii Nauk, SSSR}, 96:469--472, 1954.
\newblock English translation: On the anomalous absorption of sound near a
  second order transition point. in: Collected papers of L. D. Landau, ed. D.
  ter Haar,(Pergamon, Oxford, 1965), pp. 626-633.

\bibitem{Grm93a}
M.~Grmela.
\newblock Weakly nonlocal hydrodynamics.
\newblock {\em Physical Review E}, 47(1):351--602, 1993.

\bibitem{GrmOtt97a}
M.~Grmela and H.~C. \"Ottinger.
\newblock Dynamics and thermodynamics of complex fluids. {I}. {D}evelopment of
  a general formalism.
\newblock {\em Physical Review E}, 56(6):6620--6632, 1997.

\bibitem{Fab99a}
M.~Fabrizio.
\newblock An evolution model for the {G}inzburg-{L}andau equations.
\newblock {\em Riv. Mat. Univ. Parma}, 2(6):155--169, 1999.

\bibitem{FabAta03a}
M.~Fabrizio, B.~Lazzari, and A.~Morro.
\newblock Thermodynamics of nonlocal electromagnetism and superconductivity.
\newblock {\em Mathematical Models and Methods in Applied Sciences},
  13(7):945--969, 2003.

\bibitem{deGPro93b}
P.~G. de~Gennes.
\newblock {\em The physics of liquid crystals}.
\newblock Oxford University Press, New York, 2 edition, 1993.

\bibitem{Ful12m}
T.~F\"ul\"op.
\newblock Thermodynamics of rheology: the standard model.
\newblock unpublished.

\bibitem{Ver97b}
J.~Verh\'as.
\newblock {\em Thermodynamics and {R}heology}.
\newblock Akad\'emiai Kiad\'o and Kluwer Academic Publisher, Budapest, 1997.

\bibitem{Fer80b}
J.~D. Ferry.
\newblock {\em Viscoelastic properties of polymers}.
\newblock John Wiley and Sons, Inc., New York-Chicester-Briabane-Toronto, 3d
  edition, 1980.
\newblock 1th ed. 1960.

\bibitem{Tsc89b}
N.W. Tschoegl.
\newblock {\em The phenomenological theory of linear viscoelastic behavior: An
  introduction}.
\newblock Springer, Berlin-etc., 1989.

\bibitem{KlaEta07a}
D.~Klatt, U.~Hamhaber, P.~Asbach, J.~Braun, and I.~Sack1.
\newblock Noninvasive assessment of the rheological behavior of human organs
  using multifrequency mr elastography: a study of brain and liver
  viscoelasticity.
\newblock {\em Physics in Medizine and Biology}, 52:7281–7294, 2007.

\bibitem{HaaSlu01a}
Y.M. Haan and G.M. Sluimer.
\newblock Standard linear solid model for dynamic and time dependent behaviour
  of building materials.
\newblock {\em Heron}, 46(1):49--76, 2001.

\bibitem{MatTak93a}
K.~Matsuki and K.~Takeuchi.
\newblock Three-dimensional in situ stress determination by anelastic strain
  recovery of a rock core.
\newblock {\em Int. J. Rock Mech. Min. Sci. \& Geomech. Abstr.}, 30:1019--1022,
  1993.

\bibitem{Mat08a}
K.~Matsuki.
\newblock Anelastic strain recovery compliance of rocks and its application to
  in situ stress measurement.
\newblock {\em Int. J. Rock Mech. Min. Sci.}, 45:952--965, 2008.

\bibitem{LinEta10p}
W.~Lin, Y.~Kuwahara, T.~Satoh, N.~Shigematsu, Y.~Kitagawa, T.~Kiguchi, and
  N.~Koizumi.
\newblock A case study of 3d stress orientation determination in {S}hikoku
  island and {K}ii peninsula, japan.
\newblock In Ivan Vrkljan, editor, {\em Rock Engineering in Difficult Ground
  Conditions (Soft Rock and Karst)}, pages 277--282, London, 2010. Balkema.
\newblock Proceedings of Eurock'09 Cavtat, Croatia, 2009 X. 28-29.

\bibitem{ForSie06a}
S.~Forest and R.~Sievert.
\newblock Nonlinear microstrain theories.
\newblock {\em International Journal of Solids and Structures}, 43:7224--7245,
  2006.

\bibitem{JosPre89a}
D.~D. Joseph and L.~Preziosi.
\newblock Heat waves.
\newblock {\em Reviews of Modern Physics}, 61:41--73, 1989.

\bibitem{FulVan12a}
T.~F\"ul\"op and P.~V\'an.
\newblock Kinematic quantities of finite elastic and plastic deformations.
\newblock {\em Mathematical Methods in the Applied Sciences}, 35:1825–1841,
  2012.
\newblock arXiv:1007.2892v1.

\bibitem{NolSeg10a}
W.~Noll and B.~Seguin.
\newblock Basic concepts of thermomechanics.
\newblock {\em Journal of Elasticity}, 101:121--151, 2010.

\bibitem{VanPap10a}
P.~V\'an and C.~Papenfuss.
\newblock Thermodynamic consistency of third grade finite strain elasticity.
\newblock {\em Proceedings of the Estonian Academy of Sciences},
  59(2):126--132, 2010.
\newblock Proceedings of Nonlinear Wave Propagation Conf, Tallinn, 2009
  October.

\bibitem{FabMor03a}
M.~Fabrizio and A.~Morro.
\newblock Thermodynamics and second sound in a two-fluid model of helium {II};
  {R}evisited.
\newblock {\em Journal of Non-Equilibrium Thermodynamics}, 28:69--84, 2003.

\bibitem{AndEta98a}
McFadden G.~B. Anderson, D.~M. and A.~A. Wheeler.
\newblock Diffuse-interface methods in fluid mechanics.
\newblock {\em Annual Rev. in Fluid Mechanics}, 30:139--65, 1998.

\bibitem{God11a}
J.~D. Goddard.
\newblock A note on {E}ringen's moment balances.
\newblock {\em International Journal of Engineering Science}, 49:1486--1493,
  2011.

\bibitem{GurEta10b}
M.~E. Gurtin, E.~Fried, and L.~Anand.
\newblock {\em The mechanics and thermodynamics of continua}.
\newblock Cambridge University Press, 2010.

\end{thebibliography}

\end{document}